\documentclass[a4paper,fleqn]{cas-dc}

\usepackage[authoryear,longnamesfirst]{natbib}
\usepackage{parskip}
\usepackage{tabularx}

\begin{document}
\let\WriteBookmarks\relax
\def\floatpagepagefraction{1}
\def\textpagefraction{.001}

\shorttitle{Blockchain for circular economy}

\shortauthors{Abid et~al.}
\title [mode = title]{A Systematic Literature Review on the Use of Blockchain Technology in Transition to a Circular Economy}                      



%
\author[]{Ishmam Abid }
\fnmark[a]
\affiliation[1]{organization={Shahjalal University of Science and Technology},
    addressline={ Kumargaon}, 
    citysep={}, 
    postcode={Sylhet-3114}, 
    country={Bangladesh}}

\author[]{S.M. Zuhayer Anzum Fuad }
\fnmark[a]
\author[]{Mohammad Jabed Morshed Chowdhury}
\fnmark[b]

\ead{M.Chowdhury@latrobe.edu.au}
\author[]{Mehruba Sharmin Chowdhury }
\fnmark[a]
\author[3,4]{Md Sadek Ferdous }
\affiliation[2]{organization={La Trobe University},
    addressline={Bundoora}, 
    city={VIC},
    postcode={3086}, 
    country={Australia}}

    \affiliation[3]{organization={BRAC University},
    addressline={Dhaka},
    postcode={1206}, 
    country={Bangladesh}}
    \affiliation[4]{organization={Imperial College London},
    addressline={London},
    postcode={SW7 2AZ}, 
    country={UK}}
\cortext[cor3]{Corresponding authors}
\begin{abstract}
The circular economy has the potential to increase resource efficiency and minimize waste through the 4R framework of reducing, reusing, recycling, and recovering. Blockchain technology is currently considered a valuable aid in the transition to a circular economy. Its decentralized and tamper-resistant nature enables the construction of transparent and secure supply chain management systems, thereby improving product accountability and traceability. However, the full potential of blockchain technology in circular economy models will not be realized until a number of concerns, including scalability, interoperability, data protection, and regulatory and legal issues, are addressed. More research and stakeholder participation are required to overcome these limitations and achieve the benefits of blockchain technology in promoting a circular economy. This article presents a systematic literature review (SLR) that identified industry use cases for blockchain-driven circular economy models and offered architectures to minimize resource consumption, prices, and inefficiencies while encouraging the reuse, recycling, and recovery of end-of-life products. Three main outcomes emerged from our review of 41 documents, which included scholarly publications, Twitter-linked information, and Google results. The relationship between blockchain and the 4R framework for circular economy; discussion the terminology and various forms of blockchain and circular economy; and identification of the challenges and obstacles that blockchain technology may face in enabling a circular economy. This research shows how blockchain technology can help with the transition to a circular economy. Yet, it emphasizes the importance of additional study and stakeholder participation to overcome potential hurdles and obstacles in implementing blockchain-driven circular economy models.
\end{abstract}




\maketitle  

\section{Introduction}
The linear economy paradigm is currently one of the largest problems on Earth.
The linear ``take, make, and dispose'' economy is generating increased price volatility, supply chain risks, and growing pressures on resources \citep{EllenMac2015}. The only thing that moves from extraction to processing to assembly of the finished product are raw materials \citep{MICHELINI20172}. By using only raw materials for value creation, the difficulties and detrimental effects of the existing economic paradigm are expected to double in the next 20 years \citep{McKinsey}. As a result, researchers, policymakers, and business leaders are considering a new economic model in light of the economic losses, structural waste, supply and market risk, excessive resource use, and loss of natural systems that are reflected in this linear model \citep{EllenMac2015}. Therefore, the concept of circular economy (CE) is gaining considerable attention. A circular economy is one that is designed for the restoration and regeneration in mind to maintain the usability and worth of goods, parts, and materials at all times recognizing the differences between biological and technological cycles at times. Thus, CE can be essential for achieving environmental sustainability.

In a recent literature review, authors in \cite{lobo2021barriers} identified 24 key barriers to achieving CE. With the advent of Industry 4.0, blockchain technology (BCT) may provide a solution to some of these obstacles. The use of blockchain to enable the circular economy has yet to be demonstrated. This is an area of study, and only a small number of industries have been utilizing this. Researchers proposed numerous "R" frameworks for constructing CE, including the 3R \citep{3R}, 4R \citep{4R}, 6R \citep{6R}, and 9R \citep{9R} (explained in Section \ref{subsec:rFramework}). The approach that best illustrates how CE operates is the 4R (reduce, reuse, recycle, and recover) framework \citep{KIRCHHERR2017221}. This paper's emphasis is on the 4R framework-based, blockchain-driven CE.

Blockchain's decentralized, distributed ledger architecture, traceable, and tamper-proof characteristics can have a significant role in achieving the above 4R framework-based CE \citep{NARAYAN2020121437}. Reuse, reduce, recycle, and recovery have all been linked to the blockchain, according to a number of authors\cite{khadke2021efficient}  \cite{rejeb2022modeling}. Blockchain technology has been adopted by a number of industries, which have decreased carbon footprints, facilitated cyclical business models, enhanced performance, and simplified communication along the supply chain, thus supporting the circular economy \citep{NARAYAN2020121437, upadhyay2021blockchain}.
Authors in \citep{adams2018blockchain} discovered that tracking the possible environmental and social factors that might create health, environmental, and safety hazards is a critical application focus for Blockchain Technology.

Although blockchain has attracted a lot of interest for its potential to address CE issues, there is no recent survey, review, or systematic literature review for a 4R-based CE employing blockchain. Hence, we carried out a thorough systematic literature review on the potential uses of blockchain in the circular economy. We have studied and analyzed how researchers and businesses have used blockchain to enable reducing, reusing, recycling, and recovering for sustainable development and the limitations of the technology in corresponding aspects.

Section 2 presents the background about blockchain and circular economy. Research methodology is discussed in section 3. We have analysed our findings using resarch question in section 4. A detail discussion is presented in section 5. Finally, we concluded in section 6.

\section{Background}



In this segment, we discuss blockchain and its different aspects, the circular economy, the adaptation of the R framework, and the research perspectives on these subjects.

\subsection{Blockchain}
Initially, blockchain was proposed as a peer-to-peer electronic currency system rather than the wide variety of applications it now sees. Despite the fact that \citeauthor{haber1990time} were the first to introduce the concept of blocks connected by cryptographic chains and developed a system to prevent tampering or alteration of data recorded with timestamps, \citeauthor{nakamoto2008bitcoin} introduced hash function methods to create blocks in the chain and proposed bitcoin as a form of decentralized electronic currency \citep{9069885}. Blockchain is now much more than just an electronic money system coupled with the industrial and  \citeauthor{buterin2014next}'s invention of smart contracts. \citeauthor{viriyasitavat2019blockchain} defined blockchain as,
\begin{quote}
    \emph{``A technology that enables immutability, and integrity of data in which a record of transactions made in a system are maintained across several distributed nodes that are linked in a peer-to-peer network''}.    
\end{quote}

To add a new block to the blockchain, different consensus algorithms are used. A consensus algorithm is a way for all peers in a blockchain network to agree on the current state of the distributed ledger. The most common consensus algorithms are PoW (Proof of Work), PoS (Proof of Stake), DPoS (Delegated Proof of Stake), PBFT (Practical Byzantine Fault Tolerance), and RAFT \citep{ferdous2021survey, ferdous2020blockchain}. In a decentralized network, these algorithms improve network security and foster trust among untrusted parties.
\subsubsection{Key Features of Blockchain}
According to \citeauthor{chowdhury2019comparative}, blockchain has a number of features that make it useful in a wide range of fields. The features are discussed below.
    \begin{itemize}
        \item {\textbf{Distributed consensus on the blockchain state:}}
        One of the most important aspects of blockchain is the different kinds of consensus algorithms. These consensus algorithms allow all of the peers to arrive at an agreement regarding the present state of the blockchain in a distributed manner.
        \item {\textbf{Immutable and irreversible blockchain state:}}
        The chain state becomes immutable and irreversible when a large number of blockchain nodes participate in the distributed consensus process. In addition, the immutability of blockchain is ensured through the usage of DLT (Distributed Ledger Technology).
        \item{\textbf{Data persistence:}}
        As long as there are nodes participating in peer-to-peer networking, data kept in the blockchain persists.
        \item{\textbf{Data provenance:}}
        A transaction is a process of storing information on a blockchain. Every transaction on the blockchain is signed by a digital signature, such as public-key cryptography, in order to preserve the data's integrity and authenticity.
        \item{\textbf{Distributed data control:}}
        Blockchain stores and retrieves data via a peer-to-peer distributed ledger. As a consequence, Blockchain has no single point of failure.
        \item{\textbf{Accountability and transparency:}}
        Because any authorized participant is able to view the current state of the blockchain as well as any transaction that has taken place between participants, it ensures accountability and transparency.
        
    \end{itemize}
\subsubsection{Types of Blockchain}
According to a survey on blockchain by \citeauthor{gamage2020survey}, there are primarily two types of blockchain:
    \begin{itemize}
        \item {\textbf{Permissionless Blockchain:}} A permissionless\\ blockchain is decentralized and open by nature. This means that any peer can join in the process of determining what blocks are added to the chain without providing identifying information, and no one is responsible for controlling entry \citep{Buterin2015On}. Bitcoin \citep{nakamoto2008bitcoin} and Ethereum \citep{buterin2014next} are instances of permissionless blockchain.
        \item{\textbf{Permissioned Blockchains:}} Permissioned blockchain, as defined in \citep{LAI2018145}, is a blockchain that requires its participants' identity authentication and authorization of network access. A central authority gives each peer the right to take part in writing or reading operations on the blockchain. One of the most widely used permissioned blockchains is Corda \citep{brown2016corda} and Hyperledger Fabric \citep{androulaki2018hyperledger}.
    \end{itemize}

\subsubsection{Smart Contracts}
\begin{figure*}[h!]
    \centering
    \includegraphics[width=\textwidth]{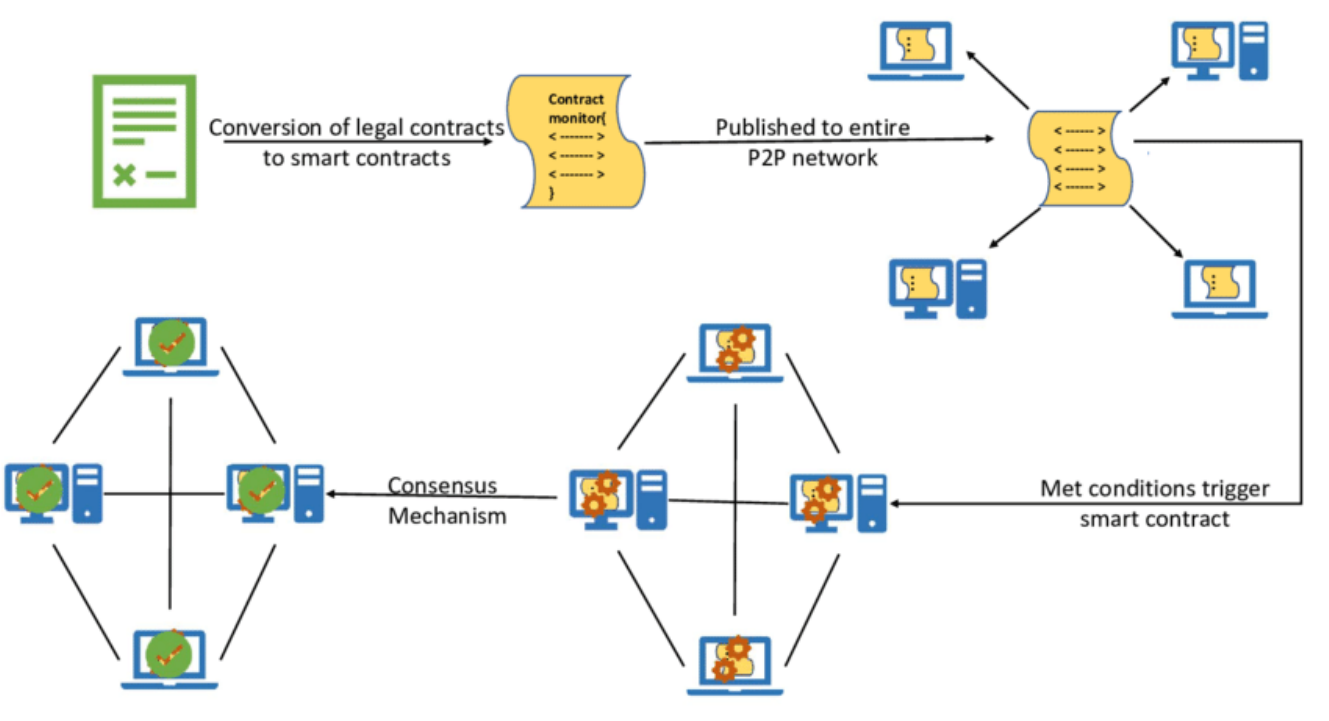}
    \caption{Smart Contract Execution Process.}
    \label{fig:1}
\end{figure*}

The idea of a smart contract has become more well-known as Blockchain 2.0 has been developed. While Szabo first introduced the idea behind smart contract \citep{szabo1997formalizing}, Vitalik Buterin later introduced Ethereum to realise the concept practically \citep{buterin2014next} in which a Turing-complete programming language was introduced for writing code and executing smart contracts in decentralized applications (dApps) on top of EVM (Ethereum Virtual Machine) and Ethereum Blockchain. a smart contract is a simple blockchain-based computer program that is executed when specific conditions are met. It is used to automate the process of putting an agreement into action so that all parties involved are aware of the outcome without the need for intermediaries or time-consuming delays \citep{IBM}. Smart contracts store data on the blockchain as transactions. This makes it possible for computer logic to be immutable. Figure \ref{fig:1} illustrates the execution processes for a smart contract.

\begin{figure*}[ht]
    \centering
    \includegraphics[width=\textwidth]{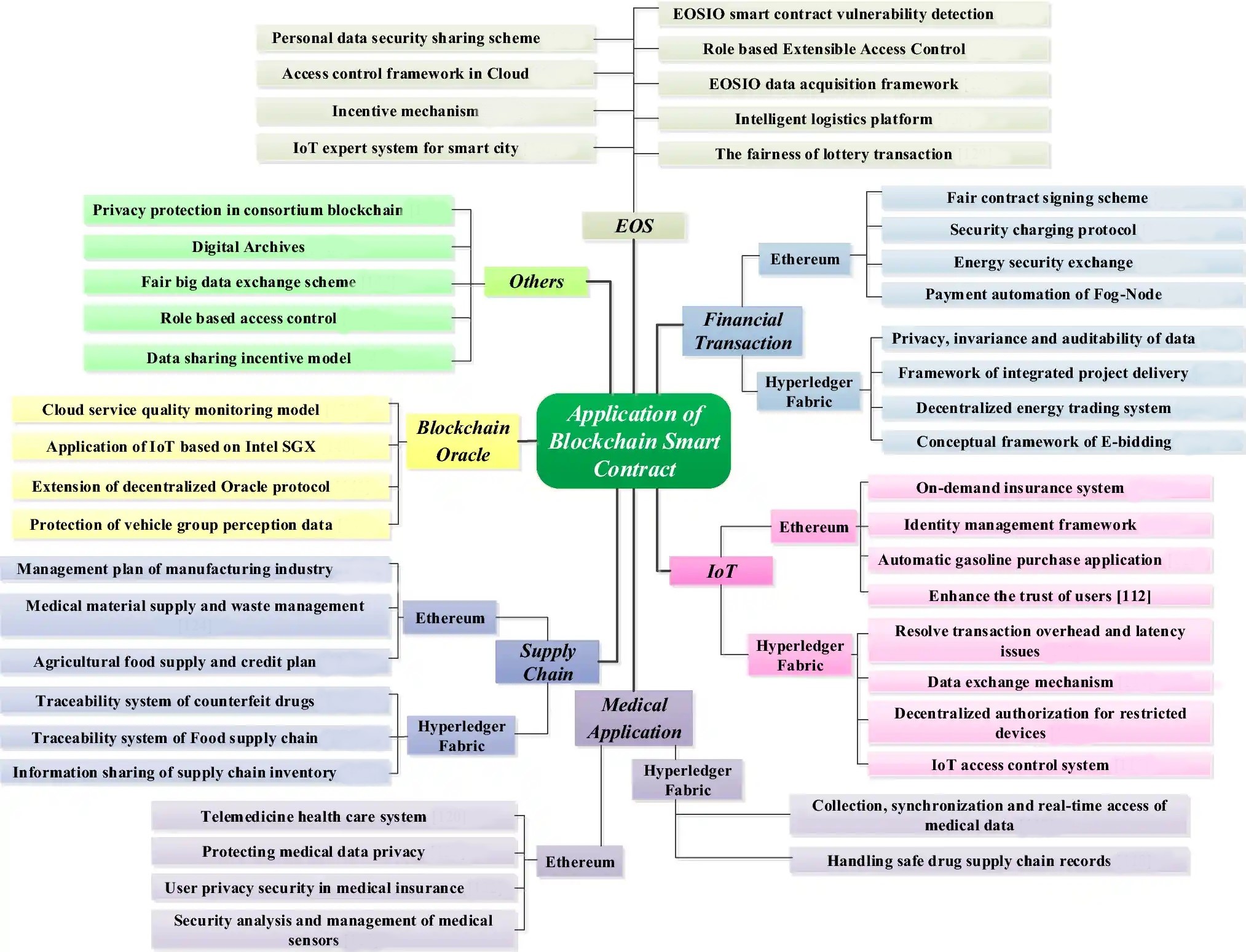}
    \caption{Application of smart contracts.\protect}
    \label{fig:2}
\end{figure*}

Smart contracts are very important for private blockchain to be functional \citep{LAI2018145}. Industry and businesses are creating smart contract-based applications to utilize blockchain in the food industry, construction industry, supply chain management, document verification, e-voting, medical data storage, FinTech, and other areas. Figure \ref{fig:2} depicts how smart contracts are widely used in the industry.

\subsection{Circular Economy and R Framework}
\sloppy
\label{subsec:rFramework}
Natural resource shortages will result from the current linear ``take-make-dispose'' economic approach. The current economic paradigm must be redesigned and shifted to a more sustainable model. Our surroundings are still being polluted by the disposal or dispersion of wastes produced by industry systems based on linear economies. A staggering 99 percent of consumer goods are thrown away within six months of purchase, which represents a dismal failure in the field of material recovery \citep{2008story}. Since the late 1970s, the notion of \textit{Circular Economy(CE)} has been gaining traction \citep{macarthur2013towards}. Several authors cite \citeauthor{pearce1990economics} as the originators of the concept. CE has been a theoretical and practical alternative to neoclassical economics since its inception. It recognizes the critical importance of the environment, its functions, and the relationship between the environment and the economic system \citep{3R}. CE defines a new approach to sustainability and social responsibility by emphasizing the social components of sustainability \citep{upadhyay2021blockchain}. \citeauthor{EllenMac2015} defines CE as below,

\begin{quote}
    \emph{``A circular economy is one that is restorative and regenerative by design and aims to keep products, components, and materials at their highest utility and value at all times, distinguishing between technical and biological cycles.''}
\end{quote}
The concept of a cyclical closed-loop and Cradle to cradle \citep{mcdonough2010cradle} systems are the common denominators across several authors who have linked the term "Circular Economy" to a wide variety of themes \citep{murray2017circular}.

According to the World Economic Forum \citep{2014Towards}, the CE will bolster net material savings; mitigate volatility and supply concerns; drive innovation and job development; enhance land productivity and soil health; and provide long-term economic resilience. Circular business models may be a subset of ``sustainable business models'' \citep{stubbs2008conceptualizing}. The study of the CE is typically divided into three themes \citep{sehnem2019circular}:   
    \begin{itemize}
        \item Innovation in technology, organization, and society \citep{1,2} .
        \item Value chains, material flows, and applications specific to certain products \citep{3}.
        \item Tools and methods for making policy \citep{4}. 
    \end{itemize}
\sloppy
According to \citep{KIRCHHERR2017221,geng2009implementing}, CE could help improve resource productivity and eco-efficiency, reform environmental management, and realize sustainable development. As a result, businesses are increasingly willing to adopt the concept of a CE in an effort to use sustainable methods in the economy \citep{Bocken}.

\begin{figure*}[h]
    \centering
    \includegraphics[width=.95\textwidth]{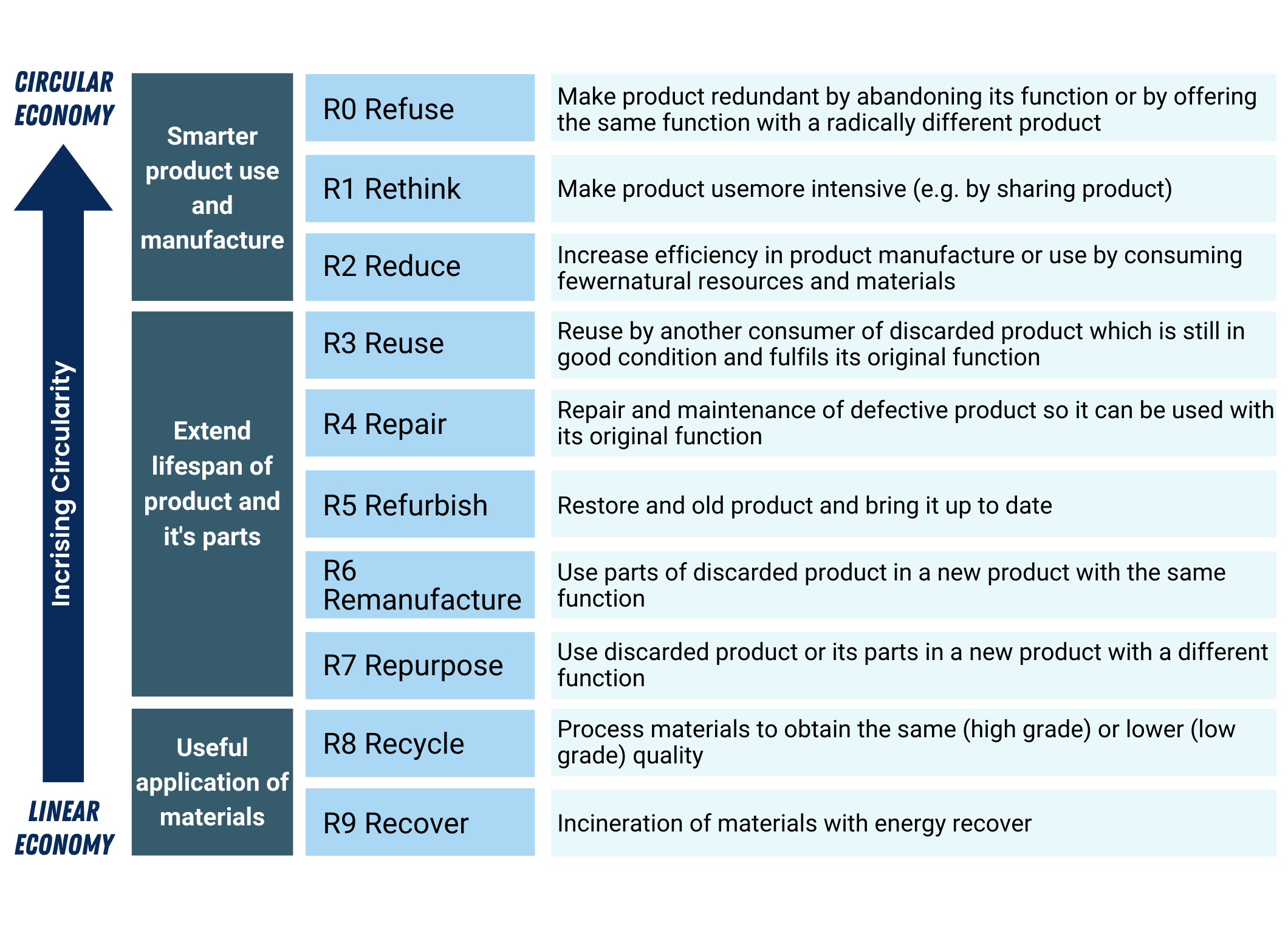}
    \caption{Brief introduction of R-strategies. Source: \protect\cite{9R}}
    \label{fig:3}
\end{figure*}

The R-framework relates to multiple techniques to embrace circularity, known as R-strategies. Authors in \citep{REIKE2018246} mentioned complex material hierarchies, also known as \textit{R-hierarchies} or \textit{R-frameworks}, as one of the key components of a more transformative perspective and evaluated and incorporated R-frameworks into a unified systemic typology comprising 10 resource value retention options (Rs) or R strategies. The 10Rs are Refuse, Rethink, Reduce, Reuse, Repair, Refurbish, Remanufacture, Repurpose, Recycle, and Recover. The majority of R-lists define a priority order for approaches to circularity, with the first R being more significant than the second R and so on \citep{9R}. Figure \ref{fig:3} illustrates a brief introduction to 10Rs.  

Researchers proposed different R-frameworks based on these R-strategies such as:
\begin{itemize}
    \sloppy
    \item \textbf{3R Framework:} Ghisellini et al. proposed 3R framework consisting of Reduce (R2), Reuse (R3), and Recycle (R8) \citep{ghisellini2016review} . The framework accounts for the circular system in which all materials are recycled and all energy is generated from renewable sources; activities support and restore the ecosystem and promote human health and a healthy society; and resources are used to create values \citep{heshmati2016review}. It is one of the most popular frameworks for achieving circularity.
    \item \textbf{4R Framework:} : Sihvonen et al. 
 introduced the 4R framework, consisting of Reduce (R2), Reuse (R3), Recycle (R8) and Recover(R9) \citep{4R}. The order of the R-strategies in the 4R framework indicates the amount of resource value retained. The higher the strategy, the more the resource value is retained \citep{henry2020typology}. The European Union’s waste framework directive is based on the 4R framework \citep{2008Directive}. The 4R Principle is one of the most prevalent principles in the field of solid waste management and sustainable development \citep{Team20204R}.
    \item \textbf{6R Framework:} Yan et al. presented the 6R approach, which consists of Reduce (R2), Reuse (R3), Recycle (R8), Recover (R9), Redesign (R7), and Remanufacture (R6) \citep{6R}. This 6R framework provides a closed-loop, multi-product life-cycle system as the basis for sustainable manufacturing \citep{jawahir2016technological}. Life Cycle Assessment (LCA) can be made more comprehensive by adding the 6R elements, and it can be used to determine the impact or burden on the environment \citep{rosenthal2016application}.
    \item \textbf{9R Framework:} Potting et al. presented the 9R framework consisting of Refuse (R0), Rethink (R1), Reduce (R2), Reuse (R3), Repair (R4), Refurbish (R5), Remanufacture (R6), Repurpose (R7), Recycle (R8) and Recover (R9) \citep{9R}. Optimizing resource and product usage is the goal of the 9R framework, which aims to create a more sustainable manufacturing capability \citep{ANG2021124264}. Using the 9R framework in advanced manufacturing, companies can achieve cleaner production and gain a competitive advantage \citep{KIRCHHERR2017221}.
\end{itemize}

These are, in conclusion, the most significant frameworks for describing CE. The aspects of these R frameworks have been accepted by the industry, which is currently seeking to devise more effective ways to implement them. There are further CE frameworks, such as ReSOLVE \citep{macarthur2015growth}. The 3R framework does not cover the entire material flow cycle, and the 6R and 9R are far too sophisticated for blockchain implementation. We utilized the 4R framework in this study due to its potential blockchain application.

\section{Research Methodology}
As part of this SLR, we reviewed prior research on how Blockchain can be utilized to establish a circular economy. Using blockchain technology, the 4Rs (Reusing, Reducing, Recycling, and Recovering) have been studied as a means of constructing a circular economy. Additionally, we examined blockchain-based industries for various CE techniques. 

\subsection{Research Questions}
After evaluating a significant number of research publications, we created five Research Questions (RQ). Table \ref{tab1} illustrates the RQs.
\begin{table}
\caption{Research questions}
\label{tab1}
\begin{tabular}{ l p{6cm} }
    \toprule
     \rowcolor{gray!30}
    \bf ID & \bf Research Questions\\
    \midrule
    \hline
    RQ1 & How blockchain technology can facilitate the \textbf{Reduce} in CE?\\
    \rowcolor{gray!15}
    RQ2 & How blockchain technology can facilitate the \textbf{Reuse} in CE?\\
    RQ3 & How blockchain technology can facilitate the \textbf{Recycle} in CE?\\
    \rowcolor{gray!15}
    RQ4 & How blockchain technology can facilitate the \textbf{Recover} in CE?\\
    RQ5 & What potential drawbacks could a blockchain-based CE pose? \\
    \bottomrule
\end{tabular}
\end{table}
\subsection{Search Strategy} 
The ultimate purpose of the search is to identify all related studies. We utilized the PRISMA-Framework for our research \citep{page2021prisma}. The inclusion-exclusion technique was adopted for archiving. We used search strings on multiple electronic databases for primary searching. We applied both forward and backward citation tracing for a secondary search. The primary selection procedure is comprised of relevant keywords, literature sources, and a screening procedure.

\subsubsection{Search terms and relevant keywords:}
\begin{figure*}[h]
    \centering
    \includegraphics[width=.95\textwidth]{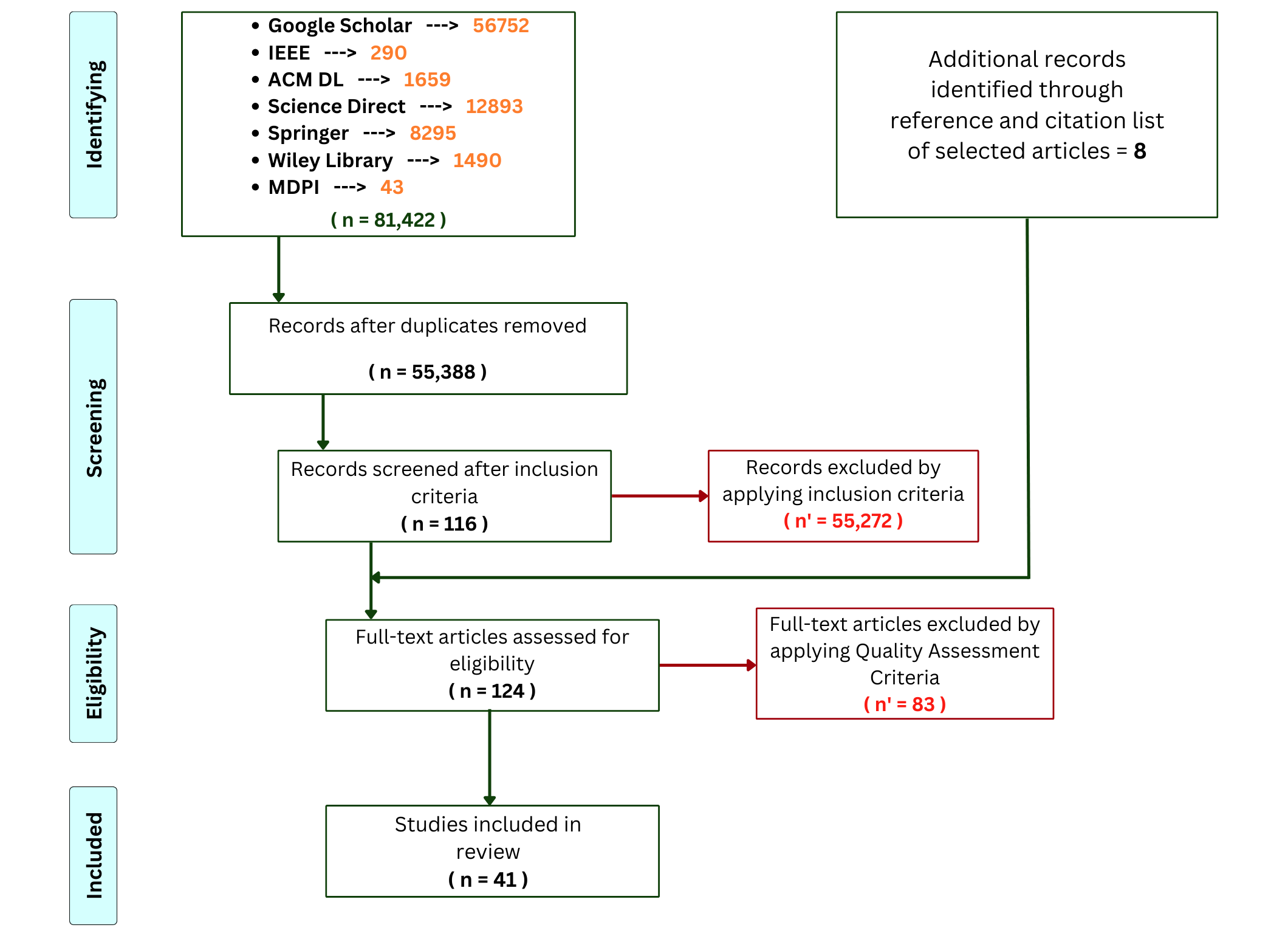}
    \caption{Applying Prisma flow diagram in our review}
    \label{fig:prisma}
\end{figure*}

We made several search strings and used them on online databases to look for studies that met our criteria.
Figure \ref{fig:prisma} shows the PRISMA Flow diagram,that depicts how different research papers are sorted for this systematic literature review. This explains how the search procedure works as a whole. Table \ref{tab2} shows the relevant keywords we used for the search.

\begin{table}[h!]
\caption{Search terms}
\label{tab2}
\begin{tabular}{ l p{5cm} }
    \toprule
     \rowcolor{gray!30}
    \bf Number & \bf Keyword\\
    \midrule
    \hline
    1 & Blockchain\\
    \rowcolor{gray!15}
    2 & Circular economy, Sustainability, Supplychain\\
    3 & 4R framework , 4R strategies \\
    \rowcolor{gray!15}
    4 & Reduce, Reuse, Recycle, Recover\\
    5 & Review, survey, SLR, literature review\\
    \bottomrule
\end{tabular}
\end{table}
\subsubsection{Literature Sources:} During this process, we used seven different electronic databases and a number of different search strings. Google Scholar, IEEE, ACM DL, ScienceDirect, Springer, Willey Online Library, and MDPI are the databases that were used. We also looked at the name of the journal, the year it was published, Bibliographies, the paper's title, the number of citations, and the link to the paper. 

\subsubsection{Search Process:}
SLR is required to conduct an exhaustive search of all sources that are relevant; hence, we have defined the search process by splitting it into two phases listed below.
\begin{itemize}
    \item \textbf{Search Term Based Searching:} Seven independent web database searches were conducted. We used search phrases from table \ref{tab2} with the logical operators "OR" and "AND," parentheses, and quote marks to narrow our search. After retrieval, the papers were included in a set of candidate papers.
    \item \textbf{Reference Based Searching:} The reference lists of the relevant papers were searched for more relevant papers and if any were discovered, they were included in the set.
\end{itemize}

As an archival of our search results, we utilized an Excel spreadsheet. We collected 81,422 articles from the original search (Table \ref{tab3}) and 8 papers from the reference search.

\begin{table}[h!]
\caption{Number of sources retrieved from the digital library.}
\label{tab3}
\begin{tabular}{ l p{4cm} }
    \toprule
     \rowcolor{gray!30}
    \bf Digital Library & \bf No of result\\
    \midrule
    \hline
    Google Scholar & 56752\\
    \rowcolor{gray!15}
    IEEE & 290\\
    ACM DL & 1659 \\
    \rowcolor{gray!15}
    ScienceDirect & 12893\\
    Springer & 8295\\
    \rowcolor{gray!15}
    Willey Online Library & 1490\\
    MDPI & 43\\
    \midrule
    \midrule
    \bf Total: &  81422 \\
    \bottomrule
\end{tabular}
\end{table}

\subsection{Selection phase}
We found a large number of candidate papers after the search process, however, not all of them were significant to our RQs. Therefore, we used an additional filtering selection process which has the following two phases:
\begin{itemize}
    \item \textbf{Inclusion-exclusion based selection:} Using an inclusion-exclusion selection strategy, we were able to pick relevant papers from our pool of candidate papers. The materials in these papers may be beneficial to our RQs. The inclusion and exclusion criteria are presented in Table \ref{tab4} and Table \ref{tab5}.
    
    \item \textbf{Final Selection:} We used some criteria to judge the quality of relevant papers during this process. After passing a quality check, certain papers were used to get the data. Section \ref{4.4} defines the standards that are employed. 
\end{itemize}
\begin{table}[h!]
\caption{Inclusion criteria for paper selection.}
\label{tab4}
\begin{tabular}{p{8cm} }
    \toprule
     \rowcolor{gray!30}
        \bf Inclusion Criteria\\
    \midrule
    \hline
    \textbullet{ } Abstracts of papers written in the English language.\\
    \rowcolor{gray!15}
     \textbullet{ } Papers published following 2012.\\
    \textbullet{ } Papers published until June 2023.\\
    \rowcolor{gray!15}\textbullet{ } Publications describing CE utilizing R-framework or R-strategies. Papers describing how blockchain can facilitate the reuse, reduction, recycling, or recovery of resources for the economy.\\
    \textbullet{ } Papers detailing the use of blockchain in various industries that attain CE.\\
    \rowcolor{gray!15}
    \textbullet{ } Papers related to the blockchain-based sustainable supply chain.\\
    \textbullet{ } If there are multiple versions of the same study, only the most recent and thorough publication is included.\\
    \rowcolor{gray!15}
    \textbullet{ } Review or survey papers on CE.\\
    \bottomrule
\end{tabular}
\end{table}
\begin{table}[h!]
\caption{Exclusion criteria for paper selection.}
\label{tab5}
\begin{tabular}{p{8cm} }
    \toprule
     \rowcolor{gray!30}
        \bf Exclusion Criteria\\
    \midrule
    \hline
    \textbullet{ } Abstracts of papers written in other languages.\\
    \rowcolor{gray!15}
     \textbullet{ } Duplicated papers found on the online database.\\
    \textbullet{} Papers that show how blockchain was used to solve problems in the traditional economic system.\\
    \rowcolor{gray!15}\textbullet{ } Discussions, comments, letters from readers, and summaries of tutorials, workshops, panels, and poster presentations.\\
    \bottomrule
\end{tabular}
\end{table}

\subsubsection{Study Quality Assessment}
\label{4.4}
We made some Quality Assessment Questions (QAQs) to make sure that the papers are of good quality. Table \ref{tab6} defines these QAQs. We chose each paper based on how many QAQs it meets in total. If a paper meets at least half of the QAQs, we have picked it in the final selection of studies. In the end, we chose 35 papers. 

\begin{table}
\caption{Research questions}
\label{tab6}
\begin{tabular}{ l p{6cm} }
    \toprule
     \rowcolor{gray!30}
    \bf ID & \bf Quality Assessment Questions\\
    \midrule
    \hline
    QAQ1 & Is that paper related to enabling CE with blockchain?\\
    \rowcolor{gray!15}
    QAQ2 & Is the solution good enough? Can it be used?\\
    QAQ3 & Does the solution fit into the R-framework or R-strategy?\\
    \rowcolor{gray!15}
    QAQ4 & Does the solution help the industry or the research community?\\
    QAQ5 & Is there a clear and thorough analysis of the limitations?\\
    \bottomrule
\end{tabular}
\end{table}



\section{Analysis}

The collected papers have been grouped into five categories: Reduce, reuse, recycle, recover, and others. 

\begin{enumerate}
    \item \textbf{Reduce} represents the studies that have discussed the use of blockchain in the transition to CE utilizing the ``Reduce'' component of the 4R-framework. This group focused primarily on the solutions provided by industries and researchers for reducing resource and material consumption using any blockchain system.
    \item \textbf{Reuse} represents the collection of studies that have discussed blockchain for the ``Reuse'' component of the 4R-framework. Specifically, these publications have used blockchain as a solution for reusing materials and resources, thus prolonging its lifespan. 
    \item \textbf{Recycle} represents the research that have been conducted on blockchain for the ``Recycle'' component of the 4R-framework. Articles which also considered the resource / waste recycling options with blockchain as the underlying technology has been placed under this category.
    \item \textbf{Recover} represents the collection of studies that have discussed blockchain for the ``Recover'' component of the 4R-framework. These researches have demonstrated how blockchain might be utilized as a ledger for waste or material recovery.
    \item \textbf{Others} represents the collection of articles that do not directly refer to any of the 4Rs but are beneficial to the community. Examples inlude resource circularity, market value development and life cycle analysis from a blockchain perspective. 
\end{enumerate}

\subsection{RQ1: How blockchain technology can facilitate Reduce in CE? }
To answer the aforementioned question, we analyzed various blockchain-based solutions proposed by other scholars that could aid in establishing the concept of ``reduce'' in CE. Numerous industry use cases aiming at eliminating obstacles linked with CE transition and the ``reduce'' idea were identified. In addition, these solutions are being discussed.
 
The notion of \textbf{reduce} in the 4R framework attempts to optimize resource consumption and reduce waste output. This can be effectively implemented through blockchain. The distributed ledger of the blockchain eliminates the need for a third-party auditor of transactions. In addition, the ability of a blockchain ledger to maintain an immutable, permanent record of transactions makes the supply chain transparent and traceable. A transparent and traceable supply chain is more significant and circular because it reduces waste and resource consumption.
\begin{itemize}
    \item {\textbf{To reduce waste: }}{
To reduce dangerous e-waste, authors in \citep{chen2021zero} have proposed analyzing the entire life cycle of electronic devices using a blockchain-based architecture. Chidepatil et al. \citep{chidepatil2020trash} demonstrated that using artificial intelligence and multi-sensor data fusion, blockchain smart contracts can help us reduce plastic waste. Using IBM's hyperledger fabric, Walmart tracked pork and mangoes along the supply chain to ensure complete traceability and reduce food waste associated with those products \citep{kouhizadeh2020blockchain}}. Kamilaris et al. \citep{kamilaris2019rise} mentioned Plastic Bank \citep{steenmans2018rubbish}, a Canadian recycling company, which was aiming to reduce plastic waste in developing countries like Haiti, Peru, Colombia, Indonesia, and the Philippines. Blockchain-secured digital tokens are given to customers who bring plastic waste to bank recycling facilities. These tokens can be used by users of the Plastic Bank app to buy additional goods. With 1 million participants, 2000 collector units around 3 million kilograms of plastic waste collected since 2014. Authors in \citep{kassou2021blockchain} suggested a blockchain-based system to reduce medical waste. They have developed a design for a blockchain-based medical and water waste management system. Users will receive digital tokens as rewards that can be exchanged for different benefits. There are Echchain, ElectricChain, Suncontract, and other platforms that use blockchain technology to reduce waste in the supply chain \citep{20177}.

\item{\textbf{To reduce intermediaries \& costs: }}Blockchain technology has an impact on administrative control and digital regulations. Data is stored in shared databases in blockchain, where it is more transparent, less likely to be deleted or changed, and immutable \citep{esmaeilian2020blockchain}. The blockchain's transparent transaction system reduces the need for intermediaries like brokers, exchanges, and banks \citep{crosby2016blockchain}. This mitigates the possibility of opportunistic behavior \citep{Saberi2018Blockchain}. According to the analysis in \citep{kouhizadeh2019nexus} analysis, the advantages of decentralization are increased by connecting buyers and sellers directly and reducing transaction costs, which can stimulate secondary market activities. Users can exchange their services and goods directly through a blockchain network. Blockchain technology improves capital flow by reducing transaction costs and investment risk. Energy systems built on the blockchain can use less electricity during long-distance transmissions. As a result, there would be less need for energy use, which would save resources and transaction costs on the network \citep{Saberi2018Blockchain}. Tushar et al. \citeauthor{tushar2020peer} highlighted the benefits of using a peer-to-peer approach to reduce the costs of energy expenditure. Small consumers will sell their excess energy units to those who do not have enough to save money. The cost of networking is reduced due to blockchain. Numerous businesses use blockchain-based crowdfunding to support the development of new platforms \citep{esmaeilian2020blockchain}. Figure \ref{fig:4} illustrates the reduction in cost scenarios using blockchain.

\begin{figure*}[h]
    \centering
    \includegraphics[width=.95\textwidth]{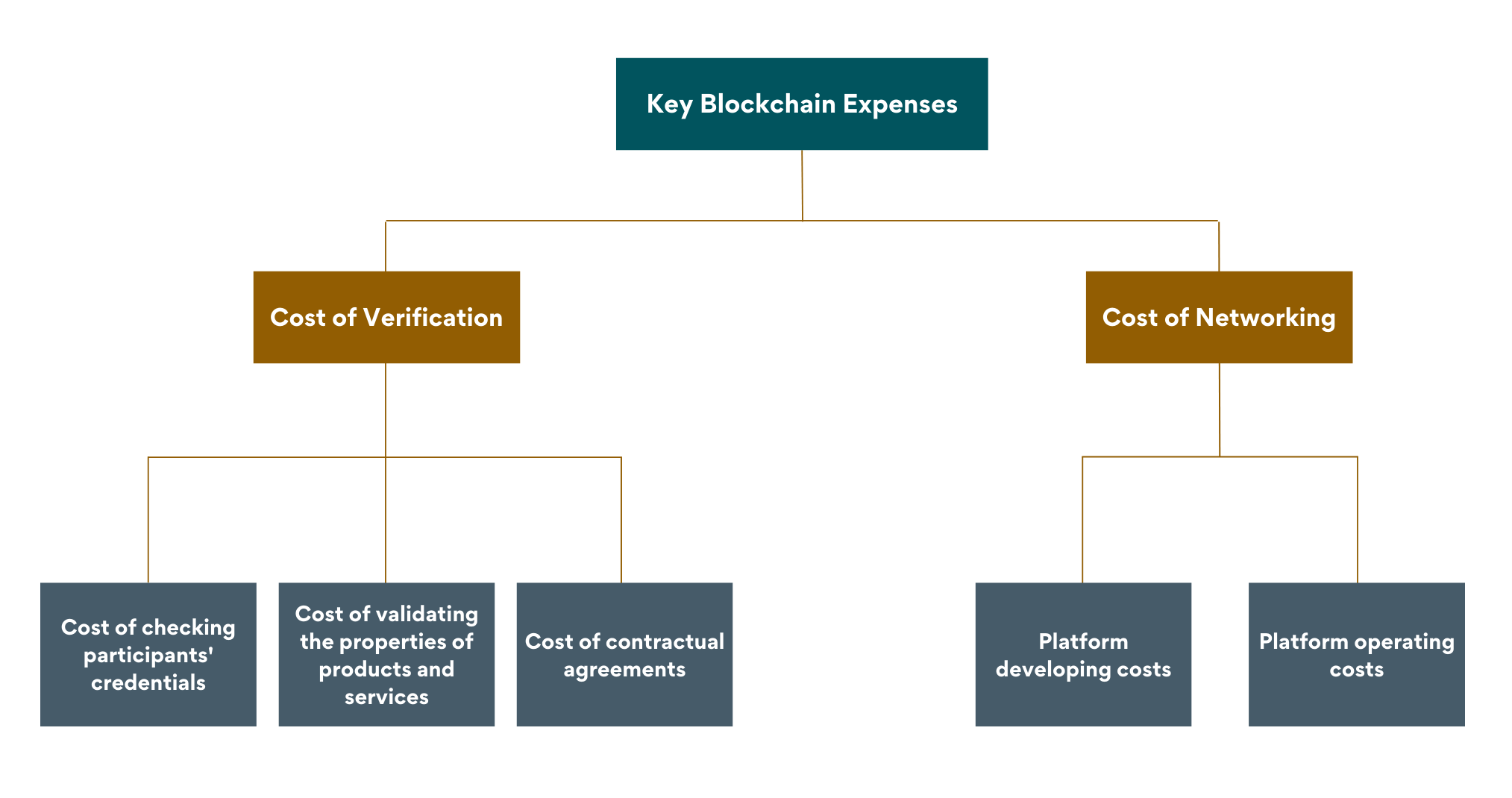}
    \caption{A brief overview of blockchain's cost-reducing potential.Source:\protect\cite{esmaeilian2020blockchain}}
    \label{fig:4}
\end{figure*}

\item{\textbf{To reduce fraud: }} Cole et al. \citep{cole2019blockchain} proposed blockchain as a solution to reduce illegal counterfeiting by disclosing a product's origin. They also mentioned the technology to reduce the cost of processes via automated systems, enabling real-time inspection through time-stamping, thereby reducing the complexity of the supply chain. Kouhizadeh et al. \citep{kouhizadeh2020blockchain} have analyzed the industry and businesses that have embraced blockchain for their products to achieve economic circularity. They cited Toyota, which utilized blockchain to reduce advertising fraud, and ad purchases. Authors in \citep{alangot2017trace} proposed developing a system to track the distribution of drugs. In this system, Internet of Things (IoT) devices such as barcode readers, smartphones, and other devices scan serial numbers or RFID tags on drug packages. They created a GDP controller that employs blockchain technology to monitor transactions and reduce fraud. In spite of the widespread consumption of genuine Australian beef, there is a substantial amount of fake beef on the market. Data61 of the CISRO (Commonwealth Scientific and Industrial Research Organisation, an Australian scientific research agency) employs blockchain technology to combat this fraud \citep{lin2017blockchain}. Blockchain improves information transparency throughout the supply chain and reduces the likelihood of data manipulation and vulnerability to crashes, fraud, and hacking \citep{zheng2018blockchain}. Reduced fraud can increase the supply chain's transparency and traceability, which benefits towards a transition to a circular economy.

\item{\textbf{To reduce overproduction: }}Blockchain technology can help in the reduction of overproduction by enabling a more effective supply chain, which will lower the consumption of raw materials and resources and speed up the transition to a circular economy. To address issues in the supply chain for the fast fashion business, \citeauthor{WANG2020103324} proposed a system based on blockchain. The three stages of a fashion item’s life — prior to production, during production, and the following production — were examined. The suggested system architecture is accessible to everyone, including fast-fashion businesses, designers, merchants, and manufacturers. Everyone must share information, keep track of inventory, and collaborate on forecasting, planning, and gap-filling for the system to function. This can make the entire process circular by reducing inventories and overproduction. The authors in \citep{9069885} presented Ethereum-based smart contracts as a means to address medical overproduction and underconsumption. The system employs four smart contracts to capture events automatically and safeguard the integrity and provenance of the data. It sets rules for the medical supply chain's phases of agreement, production, delivery, and use. Overproduction depletes resources and increases the risks for the circular economy. Xu et al \citep{Xu2019Electronics} suggested a blockchain-based system for keeping track of the electronics supply chain and identifying hardware based on certificate authorities (CA), which would make the system less vulnerable and reduce overproduction.
\end{itemize}
\subsection{RQ2: How blockchain technology can facilitate the Reuse in CE?}
Reuse is the use of discarded products, components, or materials for the same purpose for which they were originally designed, with minimum modification. When reduction is not possible, reusing is the best next option. Blockchain technology could facilitate a decentralized marketplace for reusing goods. Increasing the transparency and verifiability of information enables secondary markets for old goods and materials. Using blockchain technology, everyone can determine the quality of second-hand items \citep{shen2020selling}. On a blockchain, real-time information about reused products and resources can aid the circular economy movement. Shojaei et al. \citep{shojaei2021enabling} performed a life cycle analysis on HVAC (Heating, ventilation, and air conditioning) products, such as air conditioners, package units, gas furnaces, and split system heat pumps, using hyperledger fabric and a web interface. They kept track of the life cycles of the products to assist in decision making and proactive planning for the maximum amount of material reuse. Nandi et al. \cite{nandi2021redesigning} proposed blockchain for repairability and reuse of medical equipment. Locating devices will be used to store or deliver replacement parts. By using a public blockchain, designs for 3D-printable products will be shared for repair and reuse, also mitigating concerns over property rights. To maintain circular supply chain management in the fast fashion industry, \citeauthor{rehman2022role} have designed a system architecture and implemented blockchain for material reuse management at the application layer of the architecture. The study in \citep{Shou2022Integrating} investigated how blockchain technology can help create a secondary market for previously used leather handbags. The study delved deeper into how to keep track of used goods to facilitate the expansion of the secondary market. The results indicate that used-goods trading platforms can be established if products and their life cycles can be monitored. This would boost secondary market performance and might even make primary production obsolete. Currently, construction and demolition trash are seen as chores, although it is a consequence of the construction process. This debris can be reused and traded, and blockchain technology could be used to develop a universal waste management system that treats garbage as a resource \citep{PERERA2020100125}. According to \citep{kouhizadeh2019nexus}, blockchain technology can provide a decentralized used-goods market. With the support of information transparency and verifiability of used goods' quality and condition, the transition to a circular economy can move even more quickly and may produce new goods that utilize the secondary market. In another study, authors in \citep{electronics10162008} mentioned Cablenet, noting that the company resells its circular assets if their utility exceeds a certain threshold. This facilitates the economic reuse of products. \citeauthor{iyer2019blockchain} implemented a system to reuse wastewater utilizing IoT, machine learning, and IBM's Hyperledger fabric blockchain. Industries receive tokens in the form of cryptocurrency based on how much waste they reuse.

\subsection{RQ3: How blockchain technology can facilitate the Recycle in CE?}
\begin{figure*}[h]
    \centering
    \includegraphics[width=.95\textwidth]{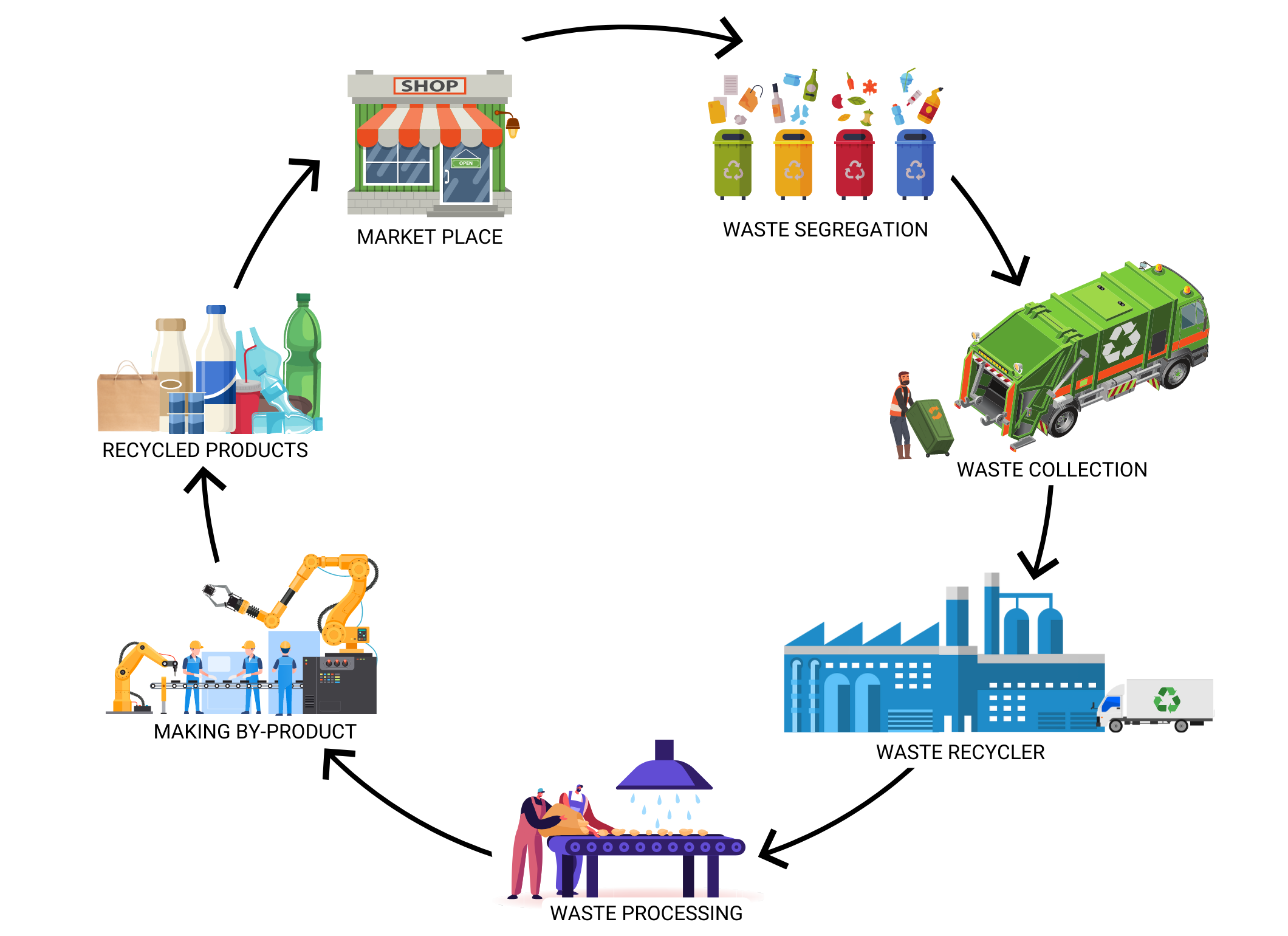}
    \caption{Recycling procedure for waste.\protect}
    \label{fig:5}
\end{figure*}
Recycling means disassembling products into their component parts and dissolving or reprocessing them into new forms. Figure \ref{fig:5} illustrates the waste recycling procedure. A blockchain-based product analysis enables the transition toward a circular business model. Using blockchain, it is possible to track all products from their point of origin through their sale and recycling. \citeauthor{centobelli2021blockchain} implemented a circular supply chain by implementing Hyperledger Fabric and a PoC (proof of concept) consensus mechanism. A circular supply network, including manufacturers, reverse logistics service providers, recycling centers, selection centers, and landfill, was modeled. Each is a participant in the permissioned blockchain. Here, the selection center collects all recyclable garbage and the recycling center recycles and distributes the items to the manufacturers. Recereum is a blockchain-based platform for profiting from trash and recyclables \citep{recereum2017recereum}. This blockchain facilitates direct communication between users and the trash collection provider. Recereum network rewards users with Recereum coins, the blockchain's native currency, based on the value of their recyclables. Ethereum powers the Recereum network. Chidepatil et al. \citep{chidepatil2020trash} introduced a blockchain-based, multi-sensor, AI-driven system for recycling plastic waste. Participants validate digital data recorded as a transaction; this data can then be used to facilitate the recycling or repurposing of plastic products. To incentivize participation in the validation process, participants receive cryptocurrency rewards. One of the major problems of the digital circular economy is motivating rivals to trade data while preserving property rights and privacy and fostering trust for recycled products \citep{ANTIKAINEN201845}. Smart contracts can be used to protect intellectual property rights and designs from counterfeiting and unauthorized usage. Cobalt Blockchain (COBC) has been offered 40,000 tons of cobalt concentrate/annum with a minimum grade of 1\% cobalt from DRC artisanal mines using blockchain to trace cobalt from mines to the point of consumption, hence enabling cobalt recycling \citep{chen2021zero}. Shojaei et al. \citeauthor{shojaei2021enabling} developed a blockchain-based system to monitor the product lifecycle throughout the supply chain. They noted that material traceability and performance records can be used to boost future output. The record of products and materials in each facility, as well as their current condition, could facilitate active recycling. The authors in \citep{kouhizadeh2019nexus} mentioned that businesses could share recyclable waste without the use of intermediaries using blockchain technology, increasing their profit margin. Furthermore, depending on aspects like the quantity, quality, and reusability of the waste, smart contracts can be used to exchange waste. Furthermore, to ensure ownership rights for waste, recyclable data can also be recorded on blockchain. Using blockchain, information about the supply chain and the recycling status of products can be stored, updated, and published. Users will then be able to identify eco-friendly products before making a purchase, promoting the circular economy \citep{wolf2022consumer}. Consumers' eco-conscious behaviors can provide a significant boost to the Circular economy transition. Numerous companies reward crypto-tokens or cryptocurrency to customers who purchase environmentally friendly products. Recycling, waste reduction, local consumption, etc. are examples of eco-friendly consumer practices \citep{esmaeilian2020blockchain}.

\subsection{RQ4: How blockchain technology can facilitate the Recover in CE?}
In the circular economy, recovery refers to the extraction of resources and compounds from waste, by-products, and residues. It is feasible to confirm the authenticity of the recovered components using blockchain technology. Once chemicals or residues have been recovered, the associated data can be recorded on a blockchain, which allows for circular economy incentives to be implemented when the recovered components are used in another product. Using smart contracts, blockchain can record the terms and conditions of waste management and related initiatives, enabling a digital waste recovery process in addition to enhancing the circular economy \citep{rejeb2022barriers}. The authors in \citep{su13094982} have provided solutions and business models to enable a circular economy, stating that blockchain is anticipated to address the inefficiencies of the traditional Extended Producer Responsibility (EPR) system by establishing a link between the product's origin and its recovery. This can accelerate the recovery process. As part of the EU Circular Foam program, Electrolux is working with polymer manufacturer Covestro to recover polyurethane foam from refrigerators \citep{MurphyHow}. Using blockchain can enable producers to record information about proper refrigerator disassembly and the most efficient method for foam recovery. Integrating blockchain with the ``Aitana'' artificial intelligence platform and Telefónica Tech's Blockchain-based TrustOs, Telefónica Tech and Exxita Be Circular created the ``Green passport'' for circular management of device life cycles \citep{Tech2022Telefonica}. The Green passport uses consumer information and device tracing mechanisms to promote device recovery. Green passports have been distributed in roughly 500,000 devices that are recovered annually. Efficient ``product return management'' will require data-driven decision-making in e-commerce reverse logistics, and blockchain application in logistics can play a significant role in value recovery \citep{dutta2020blockchain}. Blockchain can record data about material wastage and copyright and recovery processes. Consequently, any recovery facility or remanufacturer can trace products and implement recovery strategies \citep{chen2021zero}.

\subsection{RQ5: What potential barriers could a blockchain-based Circular economy face? }
Blockchain has the potential to support a circular economy, but it must first overcome obstacles such as scalability, the need for sophisticated software development tools, consumer behavior, and complex systems \citep{kouhizadeh2019nexus}. Also, organizational, financial, technological, environmental and social barriers may prevail \citep{queiroz2019blockchain}. We have researched relevant literature on the constraints of blockchain-based business models for the circular economy and have identified eight significant barriers for shifting toward a circular economy. They are: i. lack of consumers' understanding and motivation; ii. the existing linear system; iii. an expensive process of 4R products; iv. scalability and slow transactions per second (TPS); v. very few experts in blockchain; vi. inter/intra organizational obstacles; vii. government regulations and policies and viii. high resource requirements for blockchain. We have provided a brief description of each barrier in Table \ref{barrierTab} and referred to related studies that mentioned it. 
\begin{table*}[t]
  \centering
  \caption{A summary of the eight barriers with brief descriptions and references}
  \begin{tabular}{m{0.75cm} m{3.85cm} m{6cm} m{5.60cm}}
    \toprule
    \hline
    \textbf{Sr no} & \textbf{Barriers} & \textbf{Short Description} & \textbf{Related Works}\\
    \hline
    \hline
    \\
    1 & Lack of consumers' understanding and motivation & Because blockchain technology is new and circular economy activities are not prevalent, consumers have little motivation to adopt a blockchain-based circular economy. & \cite{lohmer2020blockchain, swan2015blockchain,kouhizadeh2021blockchain,ozturk2020barriers,durneva2020current,BOCKEL2021525}\\
    \\
    2 & The existing linear system &The traditional economic paradigm is ``make, use, and dispose,'' making the transition to a circular economy difficult for both customers and suppliers. & \cite{upadhyay2021blockchain, schmitz2019accounting,kouhizadeh2021blockchain}\\
    \\
    3 & The expensive processes of 4R products & Low and middle-income businesses may be unable to afford the necessary recycling equipment and chemicals to make the process profitable. & \cite{WANG2019221,wang2019digital,farooque2020fuzzy}\\
    \\
    4 & Very few experts in blockchain & Blockchain requires experts to implement the system, yet the industry lacks blockchain experts. & \cite{8720132,ozturk2020barriers,BOCKEL2021525}\\
    \\
    5 & Scalability \& Slow transactions per second (TPS) & Examples from the past indicate that scalability is a serious concern for blockchain technology. Visa, for example, has a much higher TPS than Ethereum. & \cite{swan2015blockchain,upadhyay2021blockchain,ozturk2020barriers,WANG2019221,biswas} \\
    \\
    6 & Inter/ Intra organizational obstacles & Trust issues and conflicts of interest between and within organizations can slow the progress of the blockchain-based circular economy. & \cite{galvez2018future, rehman2022role, lohmer2020blockchain,swan2015blockchain,kouhizadeh2021blockchain,ozturk2020barriers} \\
    \\
    7 & Government Regulation \& Policies & Ineffective blockchain adoption in circular economy can be attributed to a lack of corporate governance or operations administration. When developing a new strategy, the government must be consulted. Legislation is needed to deal with the risks associated with blockchain. & \cite{lohmer2020blockchain, swan2015blockchain,Saberi2018Blockchain,kamilaris2019rise,kouhizadeh2019nexus}\\
    \\
    8 & High resource requirements for blockchain & Resource-intensive consensus techniques are often used in public blockchains. PoW, as it is used in Bitcoin, requires a significant amount of computing power. &  \cite{VIRIYASITAVAT201921,swan2015blockchain,8720132,ozturk2020barriers,durneva2020current,BOCKEL2021525}\\ 
    \\
    \hline
    \bottomrule
  \end{tabular}
  
  \label{barrierTab}
\end{table*}

\section{Discussion}
The traditional ``take-make-dispose'' economic model is rapidly degrading the environment around us. Therefore, the transition to a circular economy is vital for everyone. Though blockchain is in its early stages, it already has showcased enormous potential. However, there are several obstacles that prevent it from being used to its fullest capacity. To overcome the constraints of this technology, several forces must collaborate and a massive research effort is necessary.

This study presents a comprehensive analysis of the many blockchain-based solutions that serve as foundations or drivers in the creation and implementation of CE models. In this paper, we have used research questions to investigate the potential role of blockchain technology in facilitating the transition to a circular economy within the context of the 4R framework, as well as the challenges inherent to a blockchain-based CE. We summarise our findings in this section. 

From the standpoint of our research questions, we first present the following summary. 

There were five types of categories we utilized to determine how relevant anything was to the 4Rs: (1) Reduce, (2) Reuse, (3) Recycle, (4) Recover, and (5) Applicability to all four ideas, as shown in Table \ref{refcount}.

\begin{table*}[]
\centering  
\caption{Overview of the studies/use-cases related to ``Reduce'', ``Reuse'', ``Recycle'', and ``Recover'' with Blockchain.}
\begin{tabular}{l|p{12cm}}  
  
\toprule  
\hline  
\textbf{Category} & \textbf{Related Studies/Use-cases} \\  
\hline  
\midrule 
   
 Reduce & \cite{chen2021zero}, \cite{chidepatil2020trash}, \cite{kouhizadeh2020blockchain}, \cite{kamilaris2019rise}, \cite{steenmans2018rubbish}, \cite{kassou2021blockchain}, \cite{20177}, \cite{esmaeilian2020blockchain}, \cite{crosby2016blockchain}, \cite{Saberi2018Blockchain}, \cite{kouhizadeh2019nexus}, \cite{tushar2020peer}, \cite{esmaeilian2020blockchain}, \cite{cole2019blockchain}, \cite{alangot2017trace}, \cite{lin2017blockchain}, \cite{zheng2018blockchain}, \cite{WANG2020103324}, \cite{9069885}, \cite{Xu2019Electronics}, \cite{rehman2022role}, \cite{su13169142}, \cite{saberi2019blockchain} \\\\ \hline 
 \\
 Reuse & \cite{shen2020selling}, \cite{shojaei2021enabling}, \cite{nandi2021redesigning}, \cite{rehman2022role}, \cite{Shou2022Integrating}, \cite{PERERA2020100125}, \cite{kouhizadeh2019nexus}, \cite{electronics10162008}, \cite{iyer2019blockchain} \\\\  \hline
 \\
 Recycle & \cite{centobelli2021blockchain}, \cite{recereum2017recereum}, \cite{chidepatil2020trash}, \cite{ANTIKAINEN201845}, \cite{chen2021zero}, \cite{shojaei2021enabling}, \cite{kouhizadeh2019nexus}, \cite{wolf2022consumer}, \cite{esmaeilian2020blockchain}
\cite{rejeb2022barriers}, \cite{kouhizadeh2020blockchain}, \cite{rehman2022role}, \cite{electronics10162008}, \cite{su13169142}, \cite{admsci10020023} \\\\  \hline
\\
 Recover & \cite{rejeb2022barriers}, \cite{su13094982}, \cite{MurphyHow}, \cite{Tech2022Telefonica}, \cite{dutta2020blockchain}, \cite{chen2021zero}, \cite{kouhizadeh2019nexus}, \cite{admsci10020023} \\\\  \hline
 \\
 Applicability to all four concepts & \cite{kouhizadeh2019nexus} \\\\\hline
\bottomrule  
  
\end{tabular}  
\label{refcount}
\end{table*}

The most striking finding of Table \ref{refcount} is that most of the research is primarily focused on the ``reduce'' CE notion. This fits perfectly the structured organization of the CE frameworks. Putting ``reduce'' ahead of ``reuse``, ``recycle'' and ``recover'' is crucial because it helps avoid problems such as quality degradation that can occur during recycling and reuse, the consumption of resources required by recycling and restoration, and so on. In ``reduce'' potential residual asset flows are cut off well before the product even goes into circulation. On the other hand, just a few answers were pertinent to the ``recover'' idea by itself. And unfortunately, only one research revealed applicability to all four concepts. Not only are there a few theoretical examples of how CE might profit from blockchain technology, but there are also limited actual case studies that deal with sustainability challenges. Since all of the 4R concepts are closely related to each other, finding a solution for one CE concept does not mean that it cannot also be used for another CE concept. That is why it is important to study this 4R approach in greater depth.

\begin{figure*}[h]
    \centering
    \includegraphics[width=.75\textwidth]{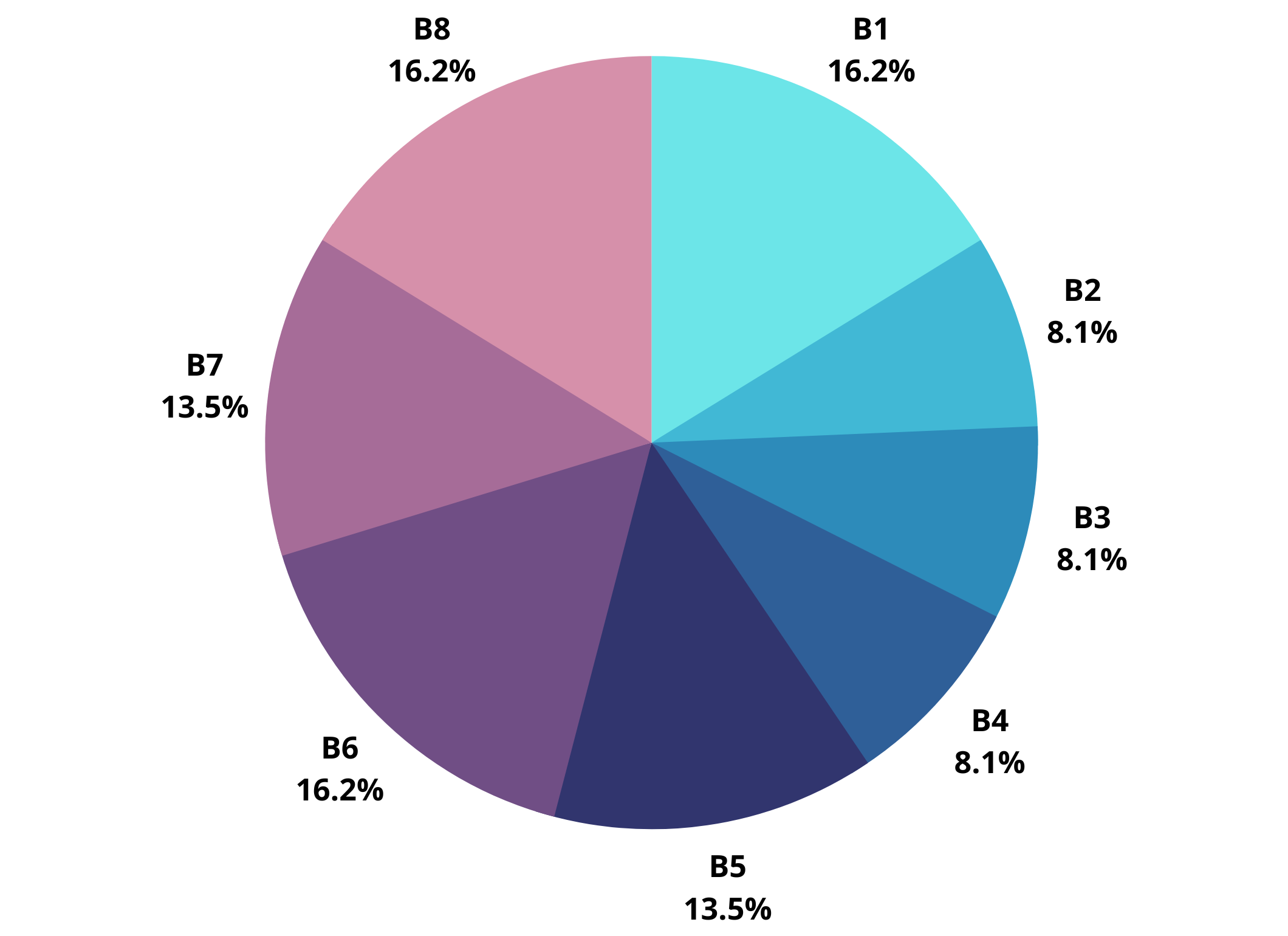}
    \caption{Percentage of frequently cited barriers in related studies/use-cases.\protect}
    \label{fig:6}
\end{figure*}

Table \ref{barrierTab} also shows that the most frequently cited barriers in the transition to a circular economy are the lack of customer knowledge and motivation, inter/intra-organizational obstacles, and high resource requirements for blockchain. Figure \ref{fig:6} presents the percentage of frequently cited barriers in the reviewed studies. 

A further significant observation is that several of the solutions utilized technology from many categories. This is because IoT, blockchain, machine learning, and AI are intertwined. Integrating blockchain with IoT in a supply chain, for instance, might boost business performance, as IoT devices are not only more productive than people but also make fewer mistakes in inventory management. In addition, it might be helpful to provide real-time traceability of goods within storage facilities, warehouses, or other places, which can help reduce product damage and maximize its utility.

\section{Conclusion and further research scopes}

There is no doubt that blockchain technology can bring benefits to the notion of circular economy. By altering the current state of recordkeeping and the value proposition, blockchain will accelerate the entire process. However, blockchain have obstacles to overcome.

This article summarizes previous research in this field and concludes that the 4R-framework of circular economy (Reduce, Reuse, Recycle, Recover) can be successfully implemented with blockchain serving as a key enabler. This article classifies blockchain's role as a CE enabler into four distinct categories. (1) promoting a circular economy by rewarding NFTs and cryptocurrencies; (2) enhancing the transparency of the product life cycle; (3) reducing operational costs and enabling efficient systems; and (4) enhancing organizational performance through data sharing. These four categories act as catalysts for the implementation of a circular economy by extending product life, decreasing resource consumption, and providing transparency for reused and recovered products. In addition, we identified eight barriers to a CE-based circular economy, including a lack of consumer understanding and motivation, the existing linear system, an expensive process for 4R products, scalability and slow transaction per second (TPS), very few blockchain experts, inter/intra-organizational obstacles, government regulations and policies, and high resource requirements for blockchain. Simply put, blockchain is a technology, but the circularity of the economy is contingent on the vision and strategies selected by businesses to govern their processes. Blockchain can be an effective option, but additional research and optimization are necessary to expand its applications.

\printcredits

\bibliographystyle{cas-model2-names}

\bibliography{main}




\end{document}